\documentclass[pre,twocolumn]{revtex4}
\usepackage{epsfig}
\usepackage{bm}
\usepackage{amsmath}
\usepackage{amssymb}

\newcommand{\f}[2]{\frac{#1}{#2}}

\begin{document}

\title{Small-scale chaotic dynamo and spontaneous breaking of local reflectional symmetry in magnetohydrodynamics}

\author{A. Bershadskii}

\affiliation{
ICAR, P.O. Box 31155, Jerusalem 91000, Israel
}

\begin{abstract}

 It is shown, using results of direct numerical simulations, that the spontaneous breaking of local reflectional symmetry (and corresponding localized kinetic and magnetic helicities) can dominate chaotic dynamics of the small-scale MHD dynamo in the Rayleigh-B{\'e}nard convection, in kinetically forced flows under strong stratification, in the near-surface solar convection, and in kinetically forced flows at large magnetic Reynolds numbers. The notions of deterministic and helical distributed chaos have been used for this purpose. The coexistence of large- and small-scale dynamo mechanisms and applications of the obtained results to quiet and active regions of the solar photosphere have been briefly discussed.

\end{abstract}

\maketitle

\section{Introduction}

   The effective {\it  large-scale} (or the domain size-scale) MHD dynamo is often related to the non-zero global (the domain averaged) kinetic helicity $\langle {\bf v} \cdot {\boldsymbol \omega} \rangle$ (where ${\bf v}$ and $ {\boldsymbol \omega}$ are the velocity and vorticity fields, and $\langle ... \rangle$ denotes average over the spatial domain). Since in the system with global reflectional symmetry the global helicity is identically zero the non-zero global helicity indicates the absence of the global reflectional symmetry. \\
   
  The magnetic energy generated by the so-called {\it small-scale} dynamos is mainly concentrated in scales considerably smaller than the domain size. Of course, in the real systems the large and small-scale dynamos can operate simultaneously (helping or suppressing each other), but in some cases one of them can dominate. In order to study the less understood small-scale dynamos, it is useful to investigate the `pure' cases when the large-scale dynamo mechanisms are suppressed. For instance, such suppressing can be related just to the global reflection symmetry of the system, which implies zero (or negligible) global helicity. \\
  
  The global reflection symmetry of a system (and, consequently, zero global helicity) does not mean that the point-wise helicity (helicity fluctuations) is also identically zero. Moreover, there can be a spontaneous breaking of the local reflectional symmetry in the chaotic/turbulent flows with the appearance of the vorticity blobs \cite{moff1}-\cite{bkt} with sign-definite (positive or negative) blob's helicity. To maintain zero global helicity such localized positive and negative blob's helicities have to be canceled at the global average. \\
  
  Taking into account the crucial role of the global helicity for the large-scale dynamos a question then arises: What role can the spontaneous breaking of the local reflectional symmetry (and corresponding localized kinetic and magnetic helicity) play in the small-scale chaotic/turbulent dynamos?  \\
  
  In Section II a chaotic dynamo driven by Rayleigh-B{\'e}nard convection is investigated in respect of the deterministic chaos properties. In Section III the spontaneous breaking of local reflectional symmetry and its relation to the localized kinetic and magnetic helicity properties are investigated in detail. In Section IV the distributed chaos notion has been introduced and applied to the velocity field for the case of the small-scale dynamo studied in Section II. In Section V the near-surface solar convection has been studied in this context. In Section VI the case of a strong stable stratification with an external kinetic forcing was considered. In Section VII small-scale chaotic dynamos with a magnetic energy spectrum corresponding to deterministic chaos and with a magnetic energy spectrum corresponding to distributed chaos dominated by magnetic helicity were investigated in the kinetically forced flows with different forcing symmetries for large magnetic Reynolds numbers. In Section VIII coexistence of the large- and small-scale dynamo mechanisms has been briefly discussed using an example.

\section{Chaotic dynamo driven by Rayleigh-B{\'e}nard convection}

\subsection{Equations and boundary/initial conditions}

In the recent paper Ref. \cite{ytc} results of the direct numerical simulations (DNS) of the small-scale dynamos
driven by Rayleigh-B{\'e}nard convection in the plane layer were reported.  The authors use the standard dimensionless equations for the velocity ${\bf v} (x,y,z)$, temperature 
$\Theta  (x,y,z)$, and magnetic ${\bf B} (x,y,z)$ fields in the Boussinesq approximation for incompressible electrically conducting fluids
  
$$  
 \frac{\partial {\bf v}}{\partial t} +{\bf v}(\nabla {\bf v}) =
-\nabla p + {\rm Pm}\nabla^2{\bf v} + {\bf F},  \eqno{(1)}
$$
$$
{\bf F} =  -{\rm Pm} [{\bf B}\times(\nabla \times {\bf B})]
+\frac{{\rm Ra}{\rm Pm^2}}{{\rm Pr}}\theta{\bf e}_z,  \eqno{(2)}
$$
$$
\frac{\partial{\bf B}}{\partial t}=
\nabla^2{\bf B}+
\nabla\times({\bf v}\times{\bf B}),  \eqno{(3)}
$$
$$
\frac{\partial\theta}{\partial t}=
\frac{{\rm Pm}}{{\rm Pr}}\nabla^2\theta
-({\bf v}\cdot\nabla)\theta, \eqno{(4)}   
$$
$$
\nabla\cdot{\bf v}=0, ~~~~~~
\nabla\cdot{\bf b}=0,   \eqno{(5-6)}
$$
where ${\bf F}$ represents the electromagnetic and buoyancy forces,
$p({\bf x},t)$ is the pressure.\\

\begin{figure} \vspace{-1.3cm}\centering
\epsfig{width=.45\textwidth,file=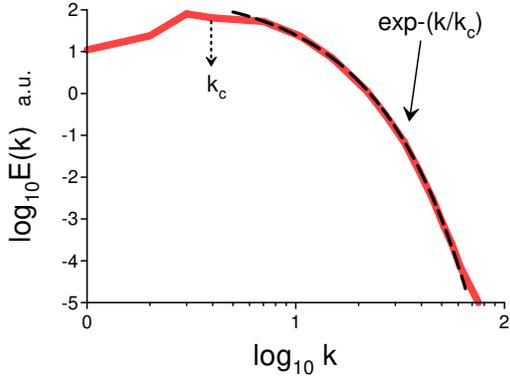} \vspace{-4.1cm}
\caption{Kinetic energy spectrum obtained for the non-magnetic Rayleigh-B{\'e}nard convection at ${\rm Ra} = 1 \times 10^4$ (and $ {\rm Pr} = 1$)} 
\end{figure}
\begin{figure} \vspace{-0.5cm}\centering
\epsfig{width=.45\textwidth,file=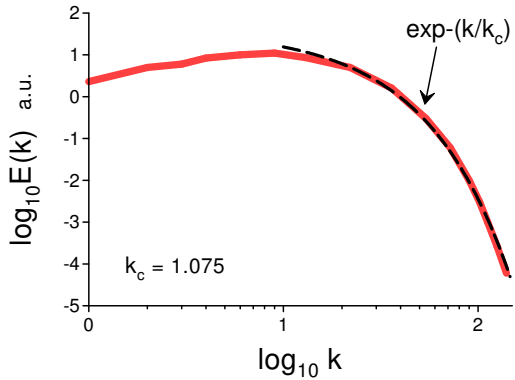} \vspace{-4.3cm}
\caption{Saturated magnetic energy spectrum obtained for the magnetic dynamo case at the same value of ${\rm Ra } = 1 \times 10^4$ (and ${\rm  Pr} = 1$) and ${\rm Pm} =5$. } 
\end{figure}
\begin{figure} \vspace{-1.6cm}\centering
\epsfig{width=.45\textwidth,file=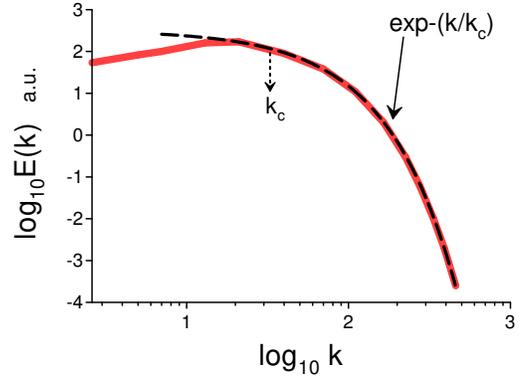} \vspace{-3.8cm}
\caption{As in Fig. 2 but for ${\rm Ra} = 4 \times 10^5$}. 
\end{figure}

 The non-dimensional parameters used in these equations:  ${\rm Ra}=\sigma g \delta T d^3/(\nu \kappa)$ denotes the Rayleigh number, $Pr=\nu/\kappa$ denotes the Prandtl number, and  $Pm=\nu/\eta$ denotes the magnetic Prandtl number. Here $g$ is the gravity acceleration (along the vertical axis $z$), $\delta T$ is the temperature difference between the plane layer boundaries, $\sigma$ is the thermal expansion coefficient, $\nu$ is the kinematic viscosity, $\kappa$ is the thermal diffusivity, $\eta$ is the magnetic diffusivity,    
 
   The boundary conditions for the velocity field:
$$
v_z = \frac{\partial v_x}{\partial z} = \frac{\partial v_y}{\partial z}  =0 ~~~~ {\rm at}~~~~ z =0,~1   \eqno{(7)}
$$
for the temperature field
$$
\theta =0~~~{\rm at}~~~ z =1,  ~~~~ {\rm and}~~~~ \theta =1 ~~~~{\rm at}~~~~ z = 1,    \eqno{(8)}
$$
and for the magnetic field
$$
B_x = B_y =0 ~~~~ {\rm at} ~~~~ z =0,~1   \eqno{(9)}
$$

It follows from the Eqs. (6) and (9) that 
$$
\frac{\partial B_z}{\partial z} = 0 ~~~~ {\rm at}~~~~ z =0,~1.   \eqno{(10)}
$$

    The dynamo simulations were started from an initial state with a rather small random magnetic field. For the values of $Ra$ larger than the critical one (for the given values of $Pm$ and $Pr$) the magnetic energy of the system undergoes a considerable growth up to a saturation state. It is important that no appreciable mean (averaged) magnetic field has been observed in the DNS. This means that the dynamo was a small-scale (fluctuating) one.\\
    
\subsection{Deterministic chaos in the non-magnetic convection}   

  The deterministic chaos was discovered in the fluid dynamics just in the case of the Rayleigh-B{\'e}nard (non-magnetic) convection \cite{lorenz}. Figure 1 shows the kinetic energy spectrum obtained for the non-magnetic case at ${\rm Ra} = 1 \times 10^4$ (and $ {\rm Pr} = 1$). The spectral data were taken from Fig. 10d of the Ref. \cite{ytc}. The dashed curve is drawn in the Fig. 1 to indicate an exponential wavenumber spectrum
 $$
 E(k) \propto \exp-(k/k_c)   \eqno{(11)}   
 $$ 
 where $k_c$ is a characteristic wavenumber. The position of parameter $k_c$ is indicated in the Fig. 1 by the dotted arrow. \\
 
  Since the exponential spectra (both spatial and temporal) are characteristic of the deterministic chaos with smooth trajectories  \cite{oh}-\cite{kds} one can conclude that here we also are dealing with the deterministic chaos. \\
   
 \subsection{Deterministic chaos in the small-scale dynamo case}
 
  Figure 2 shows the magnetic energy spectrum obtained for the full (magnetic dynamo) case at the same value of ${\rm Ra}  = 1 \times 10^4$ (and $ {\rm Pr} = 1$) and ${\rm Pm} =5$ at saturated stage. The spectral data were taken from Fig. 10b of the Ref. \cite{ytc}. The dashed curve is drawn in the Fig. 2 to indicate the exponential wavenumber spectrum Eq. (11). \\
  
  And again, from the exponential form of the spectrum, one can conclude that the magnetic field is in the state of deterministic chaos in this case. \\
  
    Figures 3 and 4 show the magnetic energy spectra obtained at the saturated stage for larger values of the parameter ${\rm Ra} =4\times 10^5,~~~{\rm and}~~~1 \times 10^7$ respectively. The exponential spectra indicate that the dynamo magnetic field is in the state of the deterministic chaos for these values of the Rayleigh number as well.\\
    
\section{Spontaneous breaking of local reflectional symmetry and helicity density moments}

\begin{figure} \vspace{-1.6cm}\centering
\epsfig{width=.47\textwidth,file=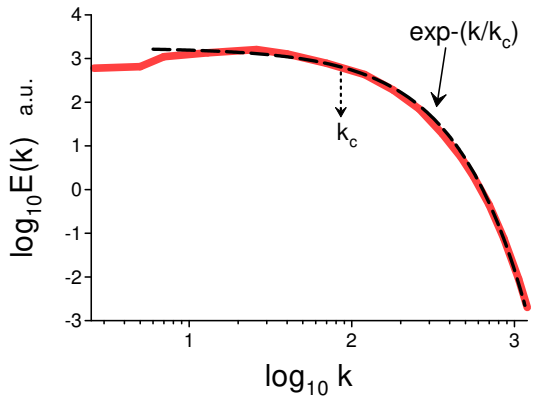} \vspace{-4.2cm}
\caption{As in Fig. 2 but for ${\rm Ra} = 1 \times 10^7 $.} 
\end{figure}
\begin{figure} \vspace{-1.45cm}\centering
\epsfig{width=.45\textwidth,file=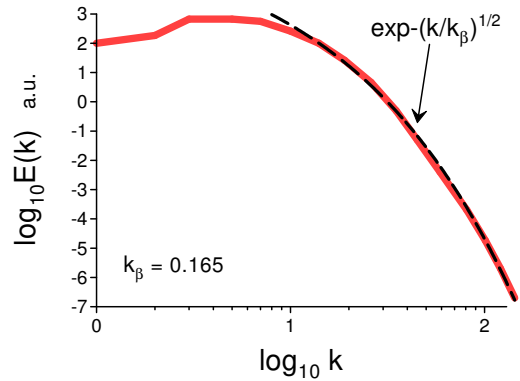} \vspace{-3.92cm}
\caption{  Saturated kinetic energy spectrum obtained for the magnetic dynamo case at the ${\rm Ra } = 1 \times 10^4$ (cf Fig. 1).} 
\end{figure}
\begin{figure} \vspace{-0.5cm}\centering
\epsfig{width=.45\textwidth,file=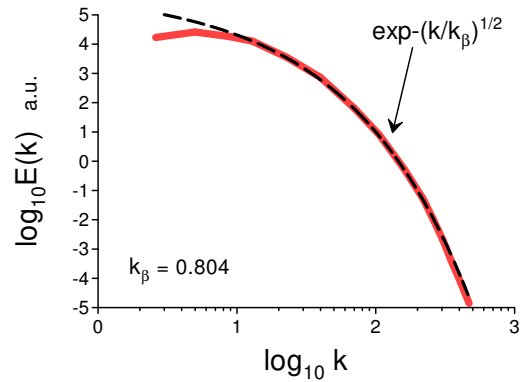} \vspace{-3.9cm}
\caption{As in the Fig. 5 but for ${\rm Ra} =4\times 10^5$. } 
\end{figure}
   The kinetic energy spectra for the dynamo case, however, are not following the simple exponential form Eq. (11) at the saturated stage of the  DNS even for the ${\rm Ra} = 1 \times 10^4$. In order to understand this let us consider the phenomenon of spontaneous breaking of local reflectional symmetry in the dynamo case.  \\
   
   As was mentioned in the Introduction there is a possibility of the appearance of the vorticity blobs moving with the fluid \cite{moff1} with considerable localized kinetic helicity even at the kinematic stage of the dynamo. The vorticity blobs contain localized kinetic helicity with opposite signs; so that the total (net) helicity of the system can be still equal to zero. Actually, the instability (generating the dynamo) can be triggered by the positive feedback through these vorticity blobs. The kinetic helicity in the vorticity blobs plays the role of a `catalyst' for the dynamo. When the magnetic field generated by the dynamo process becomes sufficiently large the magnetic blobs moving with the fluid \cite{moff1},\cite{mt} can appear. The magnetic field will be mainly concentrated in the magnetic blobs in this case, and the localized magnetic helicity in the blobs can behave similarly to the localized kinetic helicity in the vorticity blobs due to the spontaneous breaking of the local reflection symmetry. The vorticity and magnetic blobs coexist in such flows and some of them can be effectively separated from each other. This is due to the turbulent diamagnetism, over the vorticity blobs, transporting the magnetic field towards the areas of lower intensity of chaotic/turbulent vorticity and the magnetic field is concentrated between the chaotic/turbulent vorticity blobs \cite{zel}-\cite{fri}). If the chaotic/turbulent motion is localized (inhomogeneous), the associated time scale is short enough and the Lorenz (magnetic) force is not too strong, the expulsion effect from the vorticity blobs can be significant on any scale \cite{os}.\\

 Let us consider a system of the vorticity blobs moving with the fluid. The blob with number $j$ has volume $V_j$ and the vorticity is tangential at its boundary $S_j$ : ${\boldsymbol \omega} \cdot {\bf n}=0$. 
 
  The helicity in the blob $V_j$ is
$$
H_j = \int_{V_j} h({\bf r},t) ~ d{\bf r}.  \eqno{(12)}
$$

\begin{figure} \vspace{-1.5cm}\centering
\epsfig{width=.45\textwidth,file=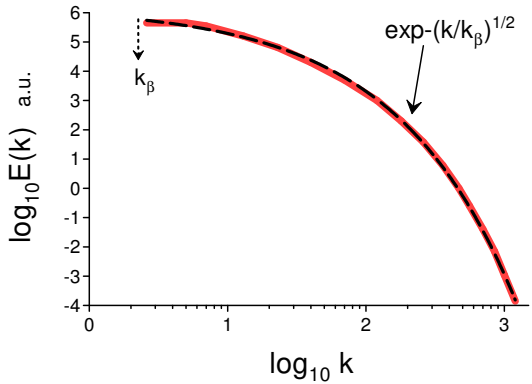} \vspace{-3.67cm}
\caption{  As in Fig. 6 but for ${\rm Ra} = 1 \times 10^7$.} 
\end{figure}

   Moments of the helicity distribution $h({\bf r},t) ={\bf v} \cdot {\boldsymbol \omega} $ can be defined as \cite{mt},\cite{lt} 
$$
{\rm I_n} = \lim_{V \rightarrow  \infty} \frac{1}{V} \sum_j H_{j}^n  \eqno{(13)}
$$
where $V$ is the total volume of the vorticity blobs.\\

 The global kinetic helicity ${\rm I_1}$ and all odd moments are identically equal to zero if the system has global reflectional symmetry. The spontaneous breaking of the local reflectional symmetry should result in the appearance of the blobs having non-zero helicity with opposite signs for different blobs; so that the global helicity and all odd moments are still identically equal to zero (or negligible). \\
 
   The global kinetic helicity ${\rm I_1}$ is generally not an invariant of the system (1-6) even for negligible dissipation due to the force ${\bf F}$ in the Eq. (1). However, the Lorenz (magnetic) force is negligible inside some vorticity blobs (due to the turbulent diamagnetism over the vorticity blobs, see above). If these vorticity blobs provide the main contribution to certain ${\rm I_n}$ Eq. (13), then the Lorenz force does not influence the dynamics of the ${\rm I_n}$.\\ 
   
   As for the buoyancy component of the force ${\bf F}$, it is concentrated mainly on the {\it large} spatial scales while just the vorticity blobs with small characteristic scales contribute mainly to the sum Eq. (13) for high moments \cite{bt}. Therefore, the high moments can be still considered as quasi-invariants of the system if the force $ {\bf F}$ Eq. (2) is mainly concentrated on large scales.\\
   
   While for the flows with the spontaneous breaking of local reflectional symmetry the {\it even} moments with the sufficiently high $n$ can be considered as finite (non-zero) quasi-invariants at the saturated stage, all odd moments have {\it zero} values due to the global reflection symmetry. \\
   
   If we denote the kinetic helicity of the vorticity blobs having negative kinetic helicity as $H_j^{-}$, and with positive helicity as $H_j^{+}$, then we can denote
$$
{\rm I_n^{\pm}} = \lim_{V \rightarrow  \infty} \frac{1}{V} \sum_j [H_{j}^{\pm}]^n  \eqno{(14)}
$$ 
with the summation in Eq. (14) for the vorticity blobs with negative (or positive) kinetic helicity only.  \\

Since for the odd moments  ${\rm I_n} = {\rm I_n^{+}} + {\rm I_n^{-}} =0$ (as a consequence of the global reflectional symmetry), then ${\rm I_n^{+}} = - {\rm I_n^{-}}$, and the odd higher moments $|{\rm I_n^{\pm}}|$ can be considered as non-zero quasi-invariants for the flows with spontaneously broken local reflectional symmetry at the saturation stage. \\

  Analogous consideration can be valid also for the above-mentioned magnetic blobs.  Let us consider a system of magnetic blobs moving with the fluid. The blob with number $j$ has volume $V_j$ and the magnetic field is tangential at its boundary $S_j$ : ${\bf B} \cdot {\bf n}=0$. \\
  
  The magnetic helicity (positive or negative) in the magnetic blob $V_j$ is
$$
^m{\rm H_j ^{\pm}}= \int_{V_j} h_m({\bf r},t) ~ d{\bf r}  \eqno{(15)}
$$
 where the magnetic helicity density $ h_m = {\bf A} \cdot {\bf B} $ and ${\bf B} = [{\nabla \times \bf A}]$. Then 
 $$
^m{\rm I_n^{\pm}} = \lim_{V \rightarrow  \infty} \frac{1}{V} \sum_j [^m{\rm H_{j}^{\pm}]^n}  \eqno{(16)}
$$ 
with the summation in Eq. (16) for the magnetic blobs with negative (or positive) magnetic helicity only.  \\
 
 The difference with the kinetic helicity is that conservation of the  $|^m{\rm I_n^{\pm}}|$ for magnetic helicity does not depend on the force ${\bf F}$ (actually it is valid for any velocity field \cite{moff1},\cite{mt}). Therefore, there is no restriction on the value of $n$ and this conservation takes place even for $n=1$ if the magnetic blobs exist in the flow due to the spontaneous breaking of local reflectional symmetry.\\

\section{Distributed chaos }

\begin{figure} \vspace{-1.5cm}\centering \hspace{-1cm}
\epsfig{width=.49\textwidth,file=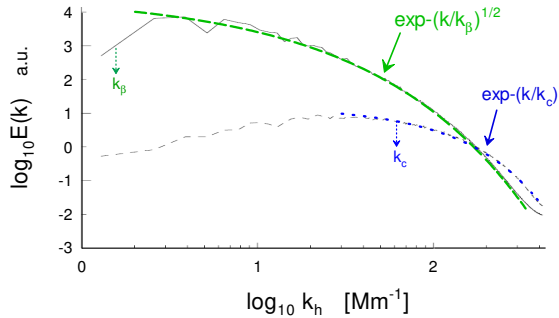} \vspace{-5.2cm}
\caption{Power spectra of the vertical components of magnetic field (lower curve) and velocity (upper curve) respectively. $k_h$ is the horizontal wave number. } 
\end{figure}
\begin{figure} \vspace{-0.45cm}\centering \hspace{-1cm}
\epsfig{width=.49\textwidth,file=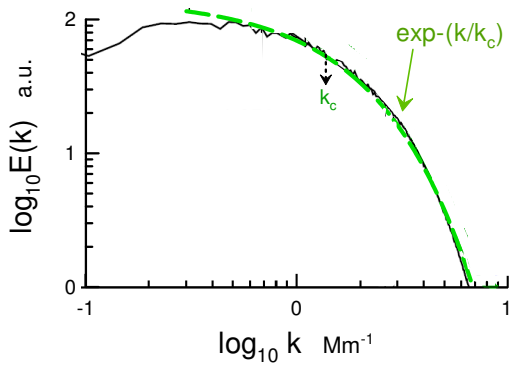} \vspace{-4.9cm}
\caption{Magnetic power spectrum computed from the magnetograms acquired by the HMI onboard SDO on June 19, 2017, over a vast quiet Sun
region at the solar disc centre. The measured magnetic field was very weak.} 
\end{figure}

   In order to evaluate the kinetic energy spectrum in the case when the characteristic scale $k_c$ in the exponential spectrum Eq. (11) fluctuates we can use an ensemble averaging
$$
E(k) \propto \int_0^{\infty} P(k_c) \exp -(k/k_c)dk_c \eqno{(17)}
$$    
with a certain probability distribution $P(k_c)$. \\

   Already the third moment can be considered as sufficiently high for quasi-invariance of $|I_3^{\pm}|$ in some cases. Of course, the moments with $n > 3$ can be also considered quasi-invariants in this case. However, the moment with a minimal value of $n$: $n_{min}$, high enough to be considered as a quasi-invariant has the maximal probability to dominate the chaotic flow, because its attractor's basin is thicker than the attractor's basins of the moments with $n > n_{min}$.\\
   
   The dimensional considerations can be used in order to relate the characteristic velocity $v_c$ to the characteristic scale $k_c$ for the chaotic flows dominated by $|I_3^{\pm}|$
$$
 v_c \propto |I_3^{\pm}|^{1/6}~ k_c^{1/2}    \eqno{(18)}
$$       
 
   Assuming Gaussian distribution (with zero mean) of the characteristic velocity $v_c$  \cite{my} and using the Eq. (18) we can readily find the probability distribution $P(k_c)$  
$$
P(k_c) \propto k_c^{-1/2} \exp-(k_c/4k_{\beta})  \eqno{(19)}
$$
with a constant parameter $k_{\beta}$.  \\

    Substituting the $P(k_c)$ from the Eq. (19) into Eq. (17) we obtain
$$
E(k) \propto \exp-(k/k_{\beta})^{1/2}  \eqno{(20)}
$$     
 
    Figures 5-7 show the saturated kinetic energy spectra obtained at the same direct numerical simulations as Figs. 2-4 for the values of the parameter ${\rm Ra} = 1 \times 10^4, ~~~{\rm Ra} =4\times 10^5,~~~{\rm and}~~~{\rm Ra} =1 \times 10^7$ respectively. The spectral data were taken from Fig. 10a of the Ref. \cite{ytc}. The dashed curves indicate correspondence to the stretched exponential spectrum Eq. (20).\\
    
    One can see that unlike the saturated magnetic field (which is in the state of deterministic chaos Figs. (2-4)) the saturated velocity field is in the state of distributed chaos Figs. (5-7) (cf Fig. 1 for the non-magnetic case).

\section{Near-surface solar convection} 

    It is known that a local near-surface small-scale convection dynamo can be operative in the quiet (undisturbed) solar photosphere (see, for instance, Ref. \cite{vs} and references therein). The quiet solar photosphere can occupy about 80\% of the solar surface and its near-surface layer is characterized by small-scale magnetic field structures and negligible global (net) helicity \cite{gra}.  \\
 
   Results of the realistic magnetohydrodynamic simulations taking into account the strong stratification (cf recent review Ref. \cite{ss}), partial ionization, compressibility, open lower boundary, and radiative transfer were reported in Ref. \cite{vs} (the authors of the Ref. \cite{vs} used the MURaM code for their computations, see for the detailed description of the code Ref. \cite{vss}). The authors reach the conclusion that non-helical, strongly stratified, compressible near-surface convection under certain conditions can work as a local self-sustained `turbulent' dynamo. Under the term `non-helical' the authors mean the absence of a global (net) helicity in their DNS (this assumption can be valid for the near-equator regions, where the Coriolis forces can be neglected). Taking into account that the consideration of Section III is valid also for compressible fluids \cite{moff1},\cite{mt} it is interesting to compare the results of the Ref. \cite{vs} with present consideration and with observational results for the quiet Sun (using present approach). \\
 
   Figure 8 shows power spectra of the vertical components of magnetic field (lower curve) and velocity (upper curve) respectively. $k_h$ is the horizontal wave number.  The spectral data were taken from Fig. 4 of the Ref. \cite{vs}. The upper dashed curve is drawn to indicate the stretched exponential spectral decay Eq. (20) for the velocity power spectrum (distributed chaos) and the lower dotted curve indicates the exponential spectral decay Eq. (11) for the magnetic field power spectrum (deterministic chaos). \\
   
   It should be noted that in this case the spontaneous breaking of the local reflection symmetry can take place already for the non-magnetic case (cf next section).\\
   
   Figure 9 shows the magnetic power spectrum computed from the magnetograms acquired by the HMI (the Helioseismic and Magnetic Imager) onboard SDO (the Solar Dynamic Observatory) on June 19, 2017. The spectral data for the Fig. 9 were taken from Fig. 2 of recent Ref. \cite{av}. The measurements were made over a vast quiet Sun
region at the solar disc centre. The measured magnetic field was very weak. The dashed curve indicates the exponential spectral decay Eq. (11) for the magnetic field power spectrum (deterministic chaos). \\

 \section{Strong stable stratification with an external kinetic forcing}
 
\begin{figure} \vspace{-0.5cm}\centering 
\epsfig{width=.47\textwidth,file=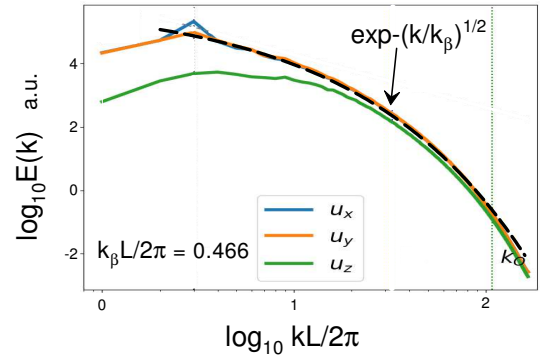} \vspace{-4.85cm} 
\caption{Kinetic energy spectrum at a kinematic stage of a small-scale dynamo in a flow generated by the kinetic external forcing under a strong stable stratification.}
\end{figure}
    
\begin{figure} \vspace{-0.45cm}\centering \hspace{-1cm}
\epsfig{width=.50\textwidth,file=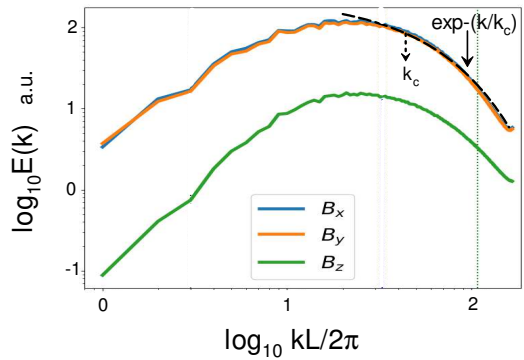} \vspace{-5.3cm}
\caption{Magnetic energy spectrum for the same flow as in the Fig. 10.} 
\end{figure}
 
   Let us now consider the kinematic stage of the small-scale dynamo at a strong stable stratification.  An interesting feature of this case is the presence of the spontaneous breaking of local reflectional symmetry in the velocity field already for non-magnetic flow. It means that despite the absence of the global helicity the vorticity blobs with localized kinetic helicity are present in the flow (due to the spontaneous breaking of local reflectional symmetry) and dominate the kinetic energy spectra already for the non-magnetic case. These localized helical structures can play a crucial role in the small-scale dynamo mechanism already at the kinematic stage.\\ 
   
   In a recent paper Ref. \cite{ssb} results of the direct numerical simulations of a small-scale dynamo in a flow generated by an isotropic, nonhelical, large-scale external kinetic forcing under a strong stable stratification were reported. \\

     The DNS was performed using the Boussinesq approximation 
$$
\partial_t \textbf{v}+\textbf{v}\cdot\nabla \textbf{v}=-\nabla p-N^2\theta {\bf e}_z+\textbf{B}\cdot\nabla \textbf{B}+\nu\nabla^2\textbf{v}+{\bf f}, \eqno{(21)}
$$
$$
\partial_t \theta+\textbf{v}\cdot\nabla \theta=v_z+\kappa\nabla^2\theta,  \eqno{(22)}
$$
$$
\partial_t \textbf{B}+\textbf{v}\cdot\nabla \textbf{B}=\textbf{B}\cdot\nabla\textbf{v}+\eta\nabla^2\textbf{B}, \eqno{(23)}
$$
$$
\nabla \cdot \textbf{v}=0, \; \nabla \cdot \textbf{B}=0,  \eqno{(24)}
$$
in a cubic, triply periodic box at $Pr =1$, $Pm =9$, and the Froude number $Fr =9$. Here $\theta$ is the buoyancy variable, $N$ is the Brunt-V\"ais\"al\"a frequency.\\

 The external kinetic forcing ${\bf f} ({\bf r},t)$ was a large-scale one in the range $2.25 < k/2\pi < 3.75$. That allows applying the consideration of Section III to this case as well. \\
   
     Figure 10 shows the kinetic energy spectrum obtained in this DNS at a kinematic stage of a small-scale dynamo. The spectral data were taken from Fig. 11a of the Ref. \cite{ssb}. The vertical dotted line separates the stratification scales (the Ozmodov scale $k_O$) and the viscous scales. The dashed curve indicates correspondence to the stretched exponential spectrum Eq. (20), i.e. the distributed chaos produced by the spontaneous breaking of the local reflectional symmetry (see above). \\
     
     Figure 11 shows the corresponding magnetic energy spectrum. The spectral data were taken from Fig. 11b of the Ref. \cite{ssb}. The dashed curve indicates correspondence to the exponential spectrum Eq. (11), i.e. the deterministic chaos (see above).\\

 \section{Large magnetic Reynolds numbers}    

\begin{figure} \vspace{-1.2cm}\centering \hspace{-1cm}
\epsfig{width=.50\textwidth,file=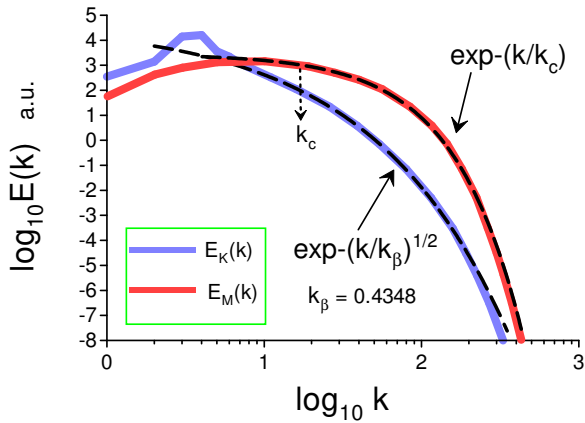} \vspace{-4.6cm} 
\caption{Kinetic and magnetic energy spectra computed in the DNS \cite{kvs} at the saturated stage for $Rm \simeq 2860$.}
\end{figure}
\begin{figure} \vspace{-0.83cm}\centering \hspace{-1cm}
\epsfig{width=.50\textwidth,file=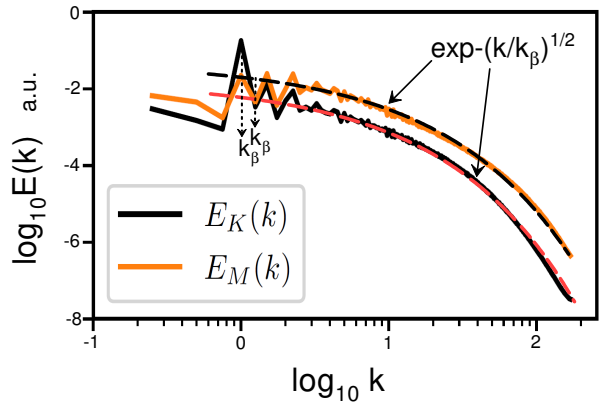} \vspace{-5cm} 
\caption{Kinetic and magnetic energy spectra computed in the DNS \cite{rin} at the saturated stage for $Rm \simeq 2800$. The wave number in the figure is normalized by $k_f$.}
\end{figure}

   The above-considered numerical simulations operated with mild magnetic Reynolds numbers $Rm \sim O(100)$. For astrophysical objects such as the Sun and galaxies the magnetic Reynolds numbers are considerably larger (for the Sun $Rm$ can be larger than $10^6$ and for the galaxies $Rm > 10^{20}$). Therefore, it is interesting to investigate the dynamo process for $Rm$ as large as possible. Moreover, it is expected that the small-scale dynamo processes can play an increasingly significant role at the large $Rm$ (see, for instance, recent Ref. \cite{rin} and references therein). They populate the turbulent plasma with localized, intense field structures which affect the system considerably faster than their large-scale dynamo counterparts. \\

   Results of direct numerical simulations with large magnetic Reynolds number $Rm \simeq 2860$  and $Pm =20$ were reported in Ref. \cite{kvs}.The kinetic forcing of the MHD dynamo equations was nonhelical (zero magnetic and kinetic helicities) and was concentrated in a narrow large-scale band $2 < k <4$ with a constant energy supply rate (buoyancy forces were not taken into account, cf Eq. (21-24)). The DNS was performed with spatially periodic boundary conditions. The simulations were started with a velocity field obtained as a result of a preliminary (pure hydrodynamical) DNS of the statistically steady isotropic homogeneous turbulence with Reynolds number $Re =666$ (the spectrum of the obtained velocity field was typical for this Reynolds number). A weak seed magnetic field was uniformly distributed over the same narrow wavenumber band as the kinetic forcing. \\
   
   Figure 12 shows kinetic and magnetic energy spectra computed in the DNS \cite{kvs} at the final time of the DNS. The spectral data were taken from Fig. 1 of the Ref. \cite{kvs}. The dashed curves indicate correspondence to the stretched exponential spectrum Eq. (20) for the kinetic energy (distributed chaos) and to the exponential spectrum Eq. (11) for the magnetic energy (deterministic chaos). Analogous results were obtained for a broadband ($ 2 < k <384$) initial seed magnetic field.\\

   Results of another DNS with large magnetic Reynolds number $Rm \simeq 2800$ and $Pm =4$ were reported in the recent Ref. \cite{rin}.  The kinetic forcing of the MHD dynamo equations was taken in the form
$$
  \begin{array}{l}
    {\bf f}({\bf r},t)=k_f\,A_f\,\times \smallskip\\  
\left(
    \begin{array}{c}
      \displaystyle{-2\sin\left(\f{2\pi y}{L_f}+\sin\omega_f t\right)\sin\f{2\pi z}{L_z}} \\
      \displaystyle{-2\cos\left(\f{2\pi x}{L_f}+\cos\omega_f t\right)\sin\f{2\pi z}{L_z}} \\
      \displaystyle{\sin\left(\f{2\pi x}{L_f}+\cos\omega_f
      t\right)+\cos\left(\f{2\pi y}{L_f}+\sin\omega_f t\right)}
      \end{array}
    \right)~ , \end{array} \eqno{(25)}
$$
where $k_f=2\pi/L_f$ is the forcing wavenumber, $L_x=L_y\equiv L_f$ and $L_z =4L_f = 4$ (buoyancy forces were not taken into account, cf Eq. (21-24)). The simulations were performed with spatially periodic boundary conditions. \\

 Despite the local helicities being rather strong in this flow the averaged over the entire spatial domain (global) helicities are zero due to the global symmetry of the forcing Eq. (25). Unlike the kinetic helicity which is not generally conserved in MHD (see Section III), the magnetic helicity is conserved in the magnetic blobs for any velocity field in the ideal case \cite{moff1},\cite{mt}. Therefore, in the case of spontaneous breaking of local reflectional symmetry the characteristic magnetic field can be estimated from the dimensional considerations using the ideal invariant $^m{\rm I_1^{\pm}}$ 
$$
 B_c \propto |^m{\rm I_1^{\pm}}|^{1/2}~ k_c^{1/2}    \eqno{(26)}
$$       
cf. Eq. (18) for the characteristic velocity. 

\begin{figure} \vspace{-1.5cm}\centering \hspace{-1cm}
\epsfig{width=.47\textwidth,file=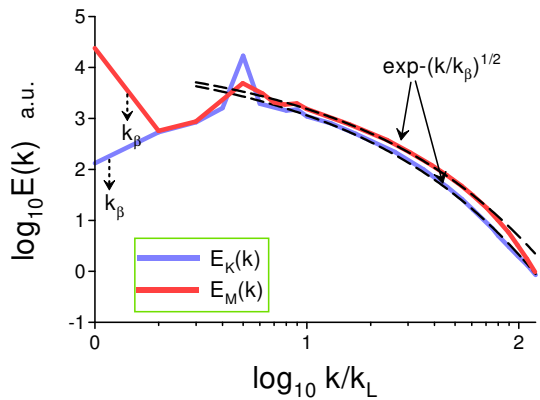} \vspace{-4.3cm} 
\caption{Kinetic and magnetic energy spectra computed in the helical DNS \cite{cb} at the saturated stage for $Rm \simeq 740$, $Pm =1$. }
\end{figure}

   If there exists a considerable global magnetic helicity it can be used in the estimate (26) instead of the  $^m{\rm I_1^{\pm}}$ with the same result.\\

   Assuming the Gaussian distribution of the characteristic magnetic field $B_c$ and using the Eq. (26) we can readily find the probability distribution $P(k_c)$  
$$
P(k_c) \propto k_c^{-1/2} \exp-(k_c/4k_{\beta})  \eqno{(27)}
$$

   Substituting the $P(k_c)$ from the Eq. (27) into Eq. (17) we obtain
$$
E(k) \propto \exp-(k/k_{\beta})^{1/2}  \eqno{(28)}
$$     
for the magnetic energy spectrum characterizing the distributed chaos in the magnetic helicity-dominated magnetic field. Remarkably it has the same form 
 as for the kinetic helicity-dominated distributed chaos in the velocity field (cf Eq. (20)). \\ 
 
    Figure 13 shows kinetic and magnetic energy spectra computed in the DNS \cite{rin} at the saturated stage. The spectral data were taken from Fig. 5 of the Ref. \cite{rin}. The dashed curves indicate correspondence to the stretched exponential spectra Eq. (20) and (28), i.e. the distributed chaos for both velocity and magnetic fields. The position of the parameter $k_{\beta}$ for the kinetic energy spectrum coincides with the pronounced peak determined by the forcing scale $k_f$ (the wavenumbers in the Fig. 13 are normalized by $k_f$).  It means that the large-scale coherent structures in the velocity field dominate the kinetic distributed chaos. That could be expected taking into account the forcing symmetries.\\
 
 \section{Coexistence of large- and small-scale dynamo mechanisms}
 
   As was already mentioned in the Introduction in real systems the large- and small-scale dynamo mechanisms usually coexist. Therefore, it can be useful to consider such coexistence using an example.  \\
   
   In Ref. \cite{cb} results of a DNS using the compressible MHD equations with an isothermal equation of state were reported. The equations had the form
$$   
   \frac{\partial {\bf A}}{\partial t} = {\bf v}\times{\bf B} -\eta\mu_{0}{\bf J},  \eqno{(29)}
$$
$$
 \frac{D{\bf v}}{D t}  = -c_s^2\nabla \ln \rho +
\frac{1}{\rho} {\bf J} \times {\bf B} + \rho^{-1}\nabla\cdot2\nu\rho{\bf S} + {\bf f},  \eqno{(30)}
$$
$$
\frac{D\ln{\rho}}{D t} = -\nabla\cdot {\bf v},  \eqno{(31)}
$$ 
where $\bf \rho$ is the density, $S_{ij}=\frac{1}{2}(v_{i,j}+v_{j,i})-\frac{1}{3}\delta_{ij}\nabla\cdot{\bf v}$ (comma denotes partial derivative), $\nu$ is the kinematic viscosity, $\mu_0$ is the vacuum permeability, $\eta$ is the molecular magnetic diffusivity, $c_s$ is the isothermal sound speed.\\

   The forcing ${\bf f} ({\bf r},t)$ is helical and centered around a wavenumber $k_f$. It is random and $\delta$-correlated in time. Relative helicity of ${\bf f} ({\bf r},t)$ is $\langle {\bf f}\cdot \nabla \times {\bf f} \rangle/[{\bf f}_{\rm rms} \cdot (\nabla \times{\bf f)}_{\rm rms}]
=2\sigma/(1+\sigma^2)$, and $\sigma$ is a measure of the helicity intensity of the forcing. It should be noted that a normalized global kinetic helicity $\langle {\bf v} \cdot {\boldsymbol \omega} \rangle/ (k_f \langle {\bf v}^2 \rangle) \simeq 2\sigma/(1+\sigma^2)$ as well.

  The DNS was performed in a cube of the size $L$ with periodic boundary conditions.\\  

  Figure 14 shows kinetic and magnetic energy spectra computed in the DNS \cite{cb} at the saturated stage. The spectral data were taken from Fig. 9b of the Ref. \cite{cb} ($k_L \equiv 2\pi/L$, $k_f = 5k_L$,  $Pm =1$, $Rm =740$, and $\sigma = 0.2$). The dashed curves indicate correspondence to the stretched exponential spectra Eq. (20) and (28), i.e. the helical distributed chaos for both velocity and magnetic fields (cf Fig 13).\\

\begin{figure} \vspace{-0.9cm}\centering 
\epsfig{width=.47\textwidth,file=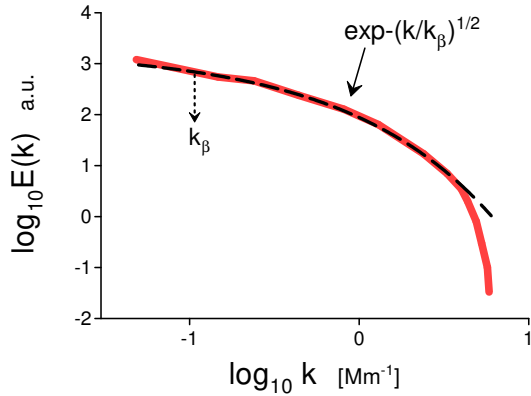} \vspace{-4.3cm} 
\caption{Magnetic power spectrum (the spot and noise corrected) of the longitudinal magnetic field in the active region NOAA 8375. }
\end{figure}

    In Section V the {\it deterministic} chaos was recognized in the dynamics of the very weak magnetic field over a vast quiet Sun region. In recent Ref. \cite{ber1} the {\it distributed} chaos was recognized in dynamics of a strong magnetic field measured in an active solar region. This distributed chaos is dominated by cross-helicity $\langle {\bf v} \cdot {\bf B} \rangle$ and has properties different from those studied in the present paper (that was determined mainly by a strong contribution of the sunspot). \\
    
   In Ref. \cite{abr} results of the SOHO/MDI (space-based) spectral measurements of the magnetic field in the active region NOAA 8375 were reported. The active region NOAA 8375 was located at the center of the solar disk at the time of the measurements (4 November 1998). The authors of the Ref. \cite{abr} blocked out the spectral contribution of the relevant sunspot. Figure 14 shows the magnetic power spectrum (the spot and noise corrected) for the measured longitudinal magnetic field. The spectral data were taken from Fig. 6b of the Ref. \cite{abr}. The dashed curve indicates the stretched exponential spectrum Eq. (28) corresponding to the distributed chaos dominated by magnetic helicity. \\

\section{Acknowledgment} 

I thank E. Levich and J. V. Shebalin for stimulating discussions.


\begin{thebibliography}{99}
\bibitem{moff1} H.K. Moffatt, J . Fluid Mech., {\bf 35}, 117 (1969)
\bibitem{lt} E. Levich and A. Tsinober, Phys. Lett. A {\bf 93}, 293 (1983)
\bibitem{mt} H.K. Moffatt and A. Tsinober, Annu. Rev. Fluid Mech., {\bf 24}, 281 (1992)
\bibitem{bkt} A. Bershadskii, E. Kit, A. Tsinober, Proc. R. Soc. Lond. A, {\bf 441}, 147 (1993)
\bibitem{oh} N. Ohtomo, K. Tokiwano, Y. Tanaka et. al., J. Phys. Soc.
Jpn., {\bf 64}, 1104 (1995)
\bibitem{sig} D.E. Sigeti, Phys. Rev. E, {\bf 52}, 2443 (1995)
\bibitem{f} J.D. Farmer, Physica D, {\bf 4}, 366 (1982).
\bibitem{fm} U. Frisch and R. Morf, Phys. Rev., {\bf 23}, 2673 (1981)
\bibitem{mm1} J. E. Maggs and G. J. Morales, Phys. Rev. Lett., {\bf 107},185003 (2011) 
\bibitem{mm2} J. E. Maggs and G. J. Morales, Phys. Rev. E {\bf 86}, 015401(R) (2012)
\bibitem{kds} S. Khurshid, D.A. Donzis and K.R. Sreenivasan, Phys. Rev. Fluids, {\bf 3}, 082601(R) (2018)
\bibitem{ytc} M. Yan, S.M. Tobias, and M.A. Calkins, , J. Fluid Mech., 915 (2021)
\bibitem{lorenz} E.N. Lorenz, J. Atmos. Sci., 20, 130 (1963)
\bibitem{zel} Ya.B. Zeldovich, JETP, {\bf 4}, 460 (1957)
\bibitem{os} M. Ossendrijver, The Astron. Astrophys. Rev., {\bf 11}, 287 (2003)
\bibitem{spen} E.J. Spence, M.D. Nornberg, C.M. Jacobson, C.A. Parada, N.Z. Taylor, R.D. Kendrick, C.B. Forest, Phys. Rev. Lett., {\bf 98}, 164503 (2007)
\bibitem{fri} P. Frick, S. Denisov, V. Noskov, A. Pavlinov, and R. Stepanov, Magnetohydrodynamics, {\bf 51}, 267 (2015)
\bibitem{bt} A. Bershadskii and A. Tsinober,  Phys. Rev. E, {\bf 48}, 282 (1993)
\bibitem{my} A. S. Monin, A. M. Yaglom, Statistical Fluid Mechanics, Vol. II: Mechanics of Turbulence (Dover Pub. NY, 2007)
\bibitem{vs} A. Voglero and M. Schussler, A\&A, {\bf 465}, L43 (2007)
\bibitem{gra} J.P. Graham, R. Cameron, and M. Schussler, ApJ., {\bf 714} 1606 (2010)
\bibitem{ss} J. Schumacher and K.R. Sreenivasan, Rev. Mod. Phys., {\bf  92}, 041001 (2020)
\bibitem{vss} A. Voglero,S. Shelyag, M. Schussler, et al., A\&A, {\bf 429}, 335 (2005)
\bibitem{av} V.I. Abramenko, and V.B. Yurchyshyn, MNRAS, {\bf 497}, 5405 (2020)
\bibitem{ssb} V. Skoutnev, J. Squire and A. Bhattacharjee, ApJ, {\bf 906}, 61 (2021)
\bibitem{kvs} R. Kumar, M.K. Verma, and R. Samtaney,  EPL, {\bf  104} 54001 (2013)
\bibitem{rin} F. Rincon, Phys. Rev. Fluids, {\bf 6}, L121701 (2021)
\bibitem{cb} S. Candelaresi and A. Brandenburg, Phys. Rev. E, {\bf 87}, 043104 (2013)
\bibitem{ber1} A. Bershadskii, Res. Notes AAS, {\bf 4}, 10 (2020)
\bibitem{abr} V. Abramenko, V. Yurchyshyn, H. Wang, P.R Goode, Solar Physics, {\bf 201}, 225 (2001)



\end{thebibliography}
\end{document}